\documentclass{article}

\usepackage{graphicx}
\usepackage[a4paper, total={7in, 10in}]{geometry}
\usepackage[square,sort,comma,numbers]{natbib}
\usepackage{subcaption}
\usepackage{color, soul}
\usepackage{xcolor}

\usepackage{authblk}

\usepackage{hyperref}

\usepackage{tikz,xcolor,hyperref}

\definecolor{lime}{HTML}{A6CE39}
\DeclareRobustCommand{\orcidicon}{%
	\begin{tikzpicture}
	\draw[lime, fill=lime] (0,0) 
	circle [radius=0.16] 
	node[white] {{\fontfamily{qag}\selectfont \tiny ID}};
	\draw[white, fill=white] (-0.0625,0.095) 
	circle [radius=0.007];
	\end{tikzpicture}
	\hspace{-2mm}
}

\foreach \x in {A, ..., Z}{%
	\expandafter\xdef\csname orcid\x\endcsname{\noexpand\href{https://orcid.org/\csname orcidauthor\x\endcsname}{\noexpand\orcidicon}}
}

\title{Plausible physical mechanisms for unusual volatile/non-volatile resistive switching in HfO$_2$-based stacks}

\author[]{C.P. Quinteros\orcidA{}}

\author[]{J. Antoja-Lleonart\orcidB{}}

\author[]{B. Noheda\orcidC{}}

\affil{\textit{Zernike Institute for Advanced Materials, University of Groningen, 9747 AG Groningen, The Netherlands}} 

\date{\today}

\begin{document}

\maketitle

\begin{abstract}
\noindent Memristive devices made of silicon compatible simple oxides are of great interest for storage and logic devices in future adaptable electronics and non-digital computing applications. A series of highly desirable properties observed in an atomic-layer-deposited hafnia-based stack, triggered our interest to investigate their suitability for technological implementations. In this paper, we report our attempts to reproduce the observed behaviour within the framework of a proposed underlying mechanism. The inability of achieving the electrical response of the original batch indicates that a key aspect in those devices has remained undetected. By comparing newly made devices with the original ones, we gather some clues on the plausible alternative mechanisms that could give rise to comparable electrical behaviours.
\end{abstract} 

\vspace{0.1cm}




\vspace{0.2cm}

\section*{Introduction}
Simple oxides are suitable candidates for microelectronic applications as they are easy to deposit and silicon compatible. Among them, hafnia-based systems are specially preferred since the material is already part of the \textit{complementary metal oxide semiconductor} (CMOS) process flow. Once interesting for its relatively high dielectric permittivity, replacing SiO$_2$ as gate oxide (to hold the capacitance while reducing the dielectric thickness), hafnia (HfO$_2$) is nowadays a subject of many research efforts by academia and industry. The discovery of ferroelectricity in doped and undoped hafnia layers \citep{boscke_ferroelectricity_2011,polakowski_ferroelectricity_2015} 
comprises a step forward towards the miniaturization needed to increase the packing density and consequently improving the overall computers' performance.  

In the meantime, a (not so) newly conceived signal processing broadly referred to as 'neuromorphic computing' \citep{schuller_neuromorphic_2015}, takes features identified in biological brains as a source of inspiration for some desired new functionalities (such as temporal coding or spiking ability). This inspiration, in addition to the striking progress made in the last decades by the \textit{Artificial Neural Networks} (\textit{ANN}), which are software implementations still running on CMOS hardware \citep{lecun_deep_2015}, seems to point towards new functionalities and huge energy savings if computer platforms include electronic devices that share some similarities with biological neurons and synapses.

The two pushes, for mimicking biological features and simplifying complex calculus operations by projecting them into the hardware, aim for more stable, more controllable, and more flexible electrical behaviours. These requirements challenge material scientists since a deep understanding of the underlying mechanisms, which in turn would allow to improve the controllability over the obtained properties, is key for unlocking some of these proposed advances.

Many requirements seem to be posed for the next generation of electronic devices \citep{schuller_neuromorphic_2015}. However, many baby steps can be done to partially solve current technological issues and set milestones in the way of a more ambitious long term goal. Aligned with that strategy, this work analyses how the properties of our devices might be promising to tackle specific issues in current designs and how to engineer them to make them suitable for a realistic technological implementation.  

This study deals with a system based on atomic-layer-deposited HfO$_2$ and sputtered TiO$_x$ on silicon and, in particular, different batches of physical samples. The original batch, which gave rise to our special interest in the stack (and whose in-depth characterization has been the subject of previous works \citep{quinteros_atomic_2018,quinteros_oxidos_2016}), is briefly introduced in subsection \textit{Preceding research} together with a description of the former microscopical picture used to rationalize the obtained electrical response. Moreover, we emphasize the reasons why such a system is remarkable both from the basic science point of view as well as the technological one. In section \textit{Results}, additional data is reported, providing new structural insight on the electronic picture of that former batch. Afterwards, we describe the attempts to reproduce the samples and the impossibility of mimicking the desired overall response. As a corollary, in the \textit{Discussion}, we consider alternative microscopical mechanisms that could produce the sought behaviour: threshold switching, ferroelectricity, and \textit{anti-ferroelectric} (\textit{AFE}) behaviour. The reconsideration of the underlying mechanism points new aspects for further research, hoping this will encourage also others to follow this approach, and finally succeed in reproducing this unique behaviour.

\subsection*{Preceding research}

The original batch was synthesized starting from a highly-doped p-type Si substrate ($\rho$ = 4-40 m$\Omega$ cm), with a thermally grown 150 nm-thick SiO$_{2-x}$\footnote{The off-stoichiometry relies on the fact that the dopants, present in the highly-doped semiconducting substrate, substantially affect the composition of the oxide achieved by this method \citep{ho_si_1979}.} layer \citep{quinteros_atomic_2018}. A metallic 20 nm-thick Ti layer was sputtered on the SiO$_{2-x}$. Subsequently, a 20nm-thick HfO$_{2}$ layer was grown by means of \textit{Atomic Layer Deposition} (\textit{ALD}), using tetrakis(dimethylamido) hafnium (TDMAHf) and ozone as hafnium and oxygen precursors, respectively \citep{quinteros_atomic_2018}. A final capping of Co (35 nm) / Pd (40 nm), photolithographically defined as square shaped electrodes of 200 $\mu$m lateral size, completed the stack. This deposition was conducted in a process that was optimized to get the most stable switching operation \citep{zazpe_thesis}.     

\begin{figure}[ht!]
        \centering
        \begin{subfigure}[b]{0.45\textwidth}
        \centering
        \includegraphics[width=\textwidth]{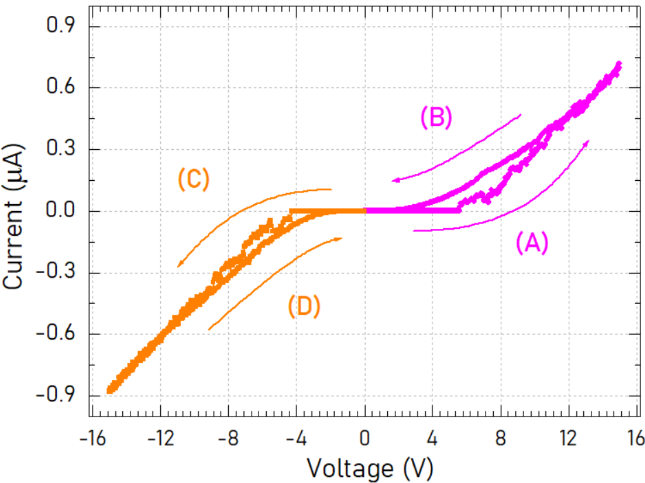}
		\subcaption{} 
		\label{fig:linIV}
        \end{subfigure}
        \hfill
        \begin{subfigure}[b]{0.47\textwidth}
        \centering
        \includegraphics[width=\textwidth]{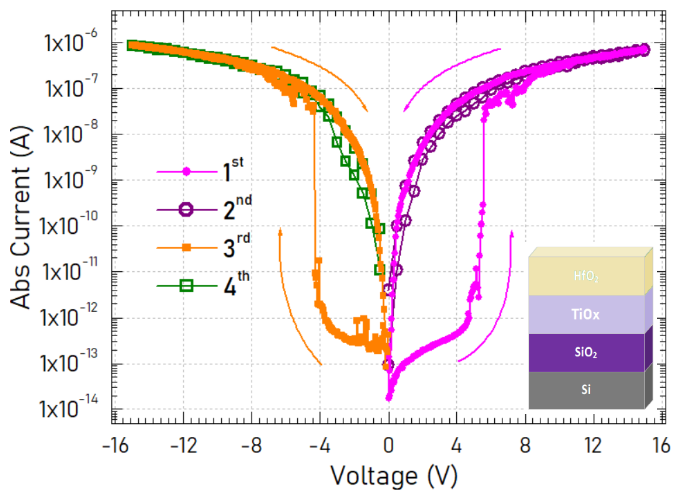}
		\subcaption{} 
		\label{fig:logIV}      
        \end{subfigure}
        \caption{Previously reported results showing a typical IV loop recorded applying a pulsed V ramp (5 ms ON / 5 ms OFF) while recording current in Auto Scale mode (adapted from \citep{quinteros_atomic_2018}). \ref{fig:linIV}) Plot in linear scale. Starting from a \textit{HRS} (A), a SET operation occurs close to + 5 V defining a \textit{LRS} that remains when reducing the stimulus amplitude (B). The two polarities appear very similar, with no measurable RESET operation but a new SET event (C), instead. Besides, \ref{fig:logIV}), plotted in semi-logarithmic scale and including four successive voltage ramps acquired in the same device, reveals the non-volatility of the achieved low-resistive states upon cycling.}
        \label{fig:IV}
\end{figure}

The devices obtained by this means displayed a remarkable IV loop (Fig. \ref{fig:IV}) that upon increasing the voltage amplitude demonstrate an abrupt change (SET) from a highly-resistive state (\textit{HRS}) to a low resistive state (\textit{LRS}) \citep{quinteros_atomic_2018}. The achieved \textit{LRS} can be retained or lost upon removing the external source depending on the specific pulsing scheme selected for ramping the voltage \citep{quinteros_atomic_2018}. Fig. \ref{fig:logIV} demonstrates that upon a 5 ms ON/5 ms OFF pulsed voltage sweep the achieved \textit{LRS}s are retained after having turned off the power supply (non-volatile). The description is qualitatively the same whether the voltage ramp goes towards positive or negative polarities. This corresponds to two switching units being present within the stack \citep{quinteros_atomic_2018}. The opposite change, from \textit{LRS} to \textit{HRS} (RESET), can be observed only if applying reading pulses in between stimulating pulses \citep{quinteros_atomic_2018}. The combination of two switching units (two SET operations observable) with the fact that each of them are bipolar, determines that upon the polarity reversal, the current is always ruled by the unit that persists in its \textit{HRS} \citep{quinteros_atomic_2018}.   

The interest in such a complex stack lies on a couple of advantageous properties that combined together make the devices very promising. Among those it is possible to mention:
\begin{itemize}
	\item \textbf{forming free}. The electrical response of the devices does not rely on a previous step to enable the switching \citep{quinteros_atomic_2018}. Upon cycling, devices behave the same as if they were pristine. 
	\item \textbf{two non-volatile bipolar units in the same stack}. One SET operation observable per polarity \citep{quinteros_atomic_2018}, similar to the so-called complementary switching \citep{linn_complementary_2010}.
	\item \textbf{rectifying behaviour}. Due to the coupling of two bipolar units, the overall response can be thought as a connection of two oppositely connected diodes \citep{quinteros_atomic_2018}.
	\item \textbf{self-limiting switching}. No external control is required to limit the runaway of the current during the SET \citep{quinteros_atomic_2018}.
    \item \textbf{high ratio current switching}. 5 orders of magnitude can be observed between the two sets of \textit{HRS} and \textit{LRS} for positive and negative polarity \citep{quinteros_atomic_2018}.
    \item \textbf{low currents for all the available states}. Even at the two \textit{LRS}: \textbf{I} $_{HRS}^{unit1}$ @+2V = 2 $\cdot$ 10$^{-13}$A,\\ \textbf{I} $_{LRS}^{unit1}$ @+2V = 10$^{-8}$A, \textbf{I} $_{HRS}^{unit2}$ @-2V = 3 $\cdot$ 10$^{-13}$A, \textbf{I} $_{LRS}^{unit2}$ @-2V = 7 $\cdot$ 10$^{-9}$A \citep{quinteros_atomic_2018}.
    \item \textbf{coexistence of volatile and non-volatile modes}. The retentivity of the states depends upon the width of and the separation in between the voltage pulses \citep{quinteros_atomic_2018}.
    \item \textbf{high repeatability along consecutive cycles} (cycle-to-cycle, \textit{C2C}, stability) \citep{quinteros_oxidos_2016}.
    \item \textbf{high reproducibility among different devices} (device-to-device, \textit{D2D}, stability) \citep{quinteros_oxidos_2016}.
\end{itemize}

However, coupled to those promising aspects, there are some others that make them unsuitable for a technological application without further optimization:

\begin{itemize}
    \item \textbf{need for (too) high voltages} (\textbf{V} $_{SET}^{unit1} \sim$ +5.5V, \textbf{V} $_{RESET}^{unit1} \sim$ -2V, \textbf{V} $_{SET}^{unit2} \sim$ -4V, \textbf{V} $_{RESET}^{unit2} \sim$ +1.5V), prohibitive for a realistic implementation \citep{quinteros_atomic_2018,quinteros_oxidos_2016}.
    \item \textbf{5-orders of magnitude ratio between two \textit{HRS} and \textit{LRS}}. This is as much an advantage, since distinguishing between the two states is quite evident, as a drawback due to the requirement for handling such dissimilar current levels \citep{quinteros_atomic_2018}.
    \item \textbf{big size electrodes} (unfeasible in terms of layout footprint) \citep{quinteros_oxidos_2016,quinteros_oxidos_2016}.
    \item \textbf{complexity of the stack} (to be reduced to its minimum) \citep{quinteros_oxidos_2016}.
    \item \textbf{too thick layers} (which would impact in the package density of any eventual device) \citep{quinteros_oxidos_2016}.
\end{itemize}

Fully understanding the physical processes that explain the observed behaviour would allow to tune the inconvenient aspects of the stacks and to explore the possibilities and limitations of the underlying physical mechanisms. The observed spatial (device-to-device) and temporal (cycle-to-cycle) stability \citep{quinteros_oxidos_2016}, lacking in many other systems, keeps pushing the effort for comprehensively understanding which is the physical mechanism that gives rise to these highly-desired properties.   

In previous works, a physical mechanism compatible with all the available experimental was proposed \citep{quinteros_oxidos_2016,quinteros_atomic_2018}. Experimental evidence of oxygen being present all the way across the Ti layer \citep{quinteros_atomic_2018} pointed out the formation of an unintentionally formed TiO$_x$ layer. The explanation, considering a stack as complex (and thick) as the one described here, was rationalized in terms of charge-trapping and the semiconducting nature of the specific TiO$_{x}$ phase \citep{quinteros_atomic_2018}. Within this model, the TiO$_{x}$ would play a key role while the SiO$_{2-x}$ and HfO$_2$ would (surprisingly) only contribute as leaky dielectrics through which electrons would flow by means of hopping between trap states. Those states would originate from the defective off-stoichiometric nature of both oxides, the SiO$_{2-x}$ due to the growth method (from a highly-doped semiconductor \citep{ho_si_1979}) and the HfO$_2$ due its known oxygen transport ability and to the oxygen scavenging of the two neighbouring layers: Ti and Co \citep{quinteros_atomic_2018}. In the present study, we report that this explanation might not be enough to account for the observed changes in the electrical response and open the discussion to other plausible underlying mechanisms. 

\section*{Results}

\subsection*{Microscopy analysis}

In the original batch (from now on referred to as batch B1), the two accessible electrodes of each device are the highly-doped Si substrate and the photo-lithographically defined Co/Pd pads. As previously mentioned, ex-situ chemical analysis has shown that oxygen is present across the whole Ti layer \citep{quinteros_atomic_2018}, featuring an unintentional oxidation of the metallic Ti layer promoted by the successive deposition steps. Consequently, the stack actually consists of a nanolaminated dielectric composed by SiO$_{2-x}$/TiO$_x$/HfO$_2$ sandwiched between two metallic-like electrodes. In order to improve our understanding of the structure, a cross-section is prepared and imaged under the electron microscope.

\begin{figure}[ht!]
        \centering
        \includegraphics[width=0.5\textwidth]{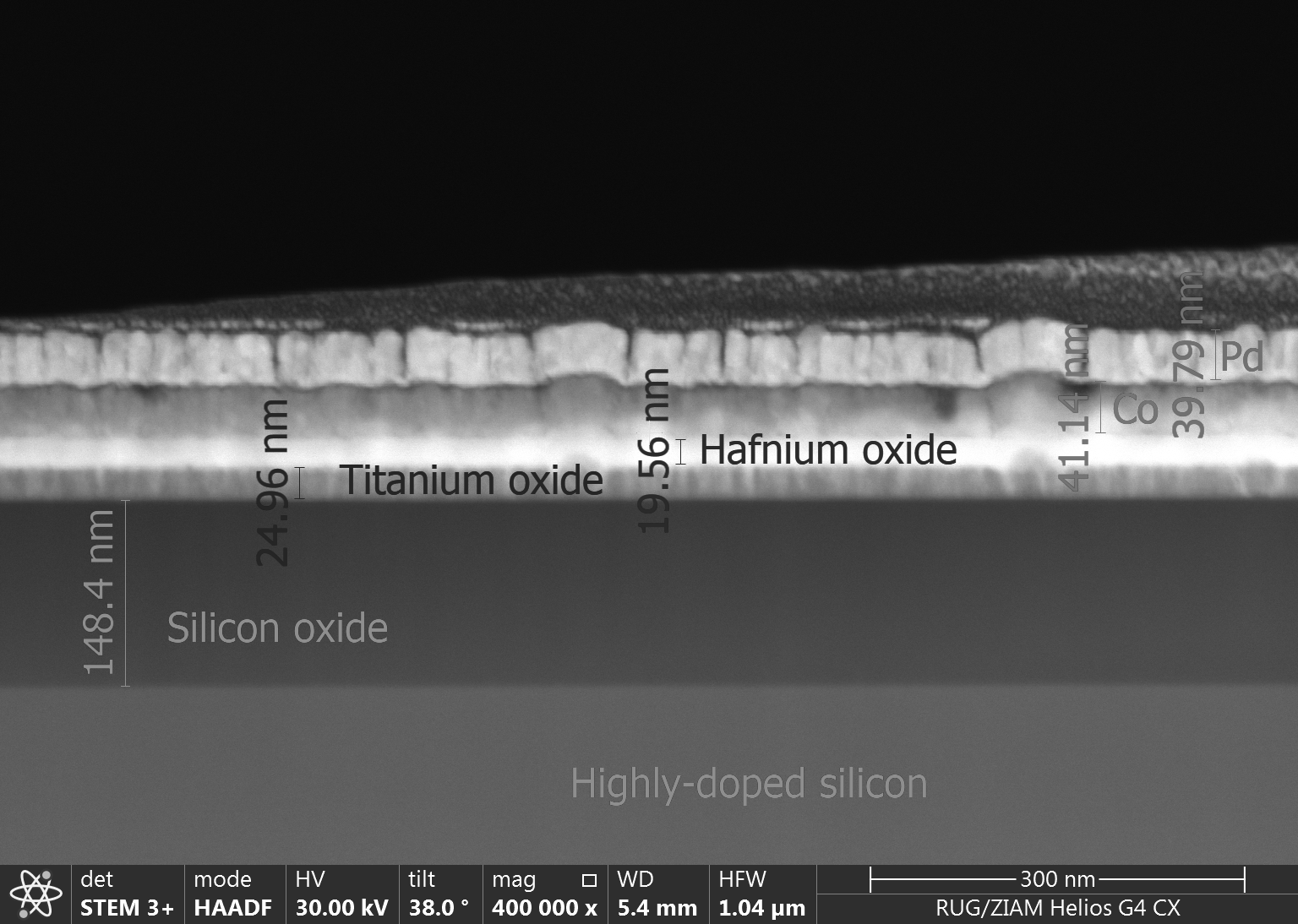}
        \caption{\textit{HAADF} image of a 1 $\mu$m-sized window of a lamella taken from a typical stack (sample B1) by means of the focus ion beam technique.}
        \label{fig:STEM}
\end{figure}

Fig. \ref{fig:STEM} shows a \textit{Scanning Transmission Electron Microscopy} (\textit{STEM}) image of a lamella prepared by means of the focus ion beam technique (Helios 660 Thermo Fisher Scientific) from one device belonging to sample B1\footnote{Through the manuscript 'samples' are understood as the pieces of substrate subjected to one specific set of deposition and processing steps. Alternatively, each top electrode defined within a sample, determines a device. Thus one sample comprises many devices.}. \textit{STEM} images are formed by convergent electrons passing through a sufficiently thin specimen ($\sim$ 100 nm) and scanned across the field of view \citep{williams_transmission_2009}. In particular, \textit{annular dark-field} (\textit{ADF}) mode was used. In \textit{ADF}, images are formed by fore-scattered electrons on an annular detector, which lies outside of the path of the directly transmitted beam. By using the \textit{high-angle ADF} (\textit{HAADF}) mode, to detect incoherently scattered electrons, it is possible to form atomic resolution images where the contrast of an atomic column is directly related to the atomic number (Z-contrast image).

Fig. \ref{fig:STEM} is a \textit{HAADF} image of a 1 $\mu$m-sized window. The stack can be easily identified (with the successive layers from bottom to top being SiO$_{2-x}$/TiO$_{x}$/HfO$_2$/Co/Pd), in turn covered by two protective layers of Pt deposited during the lamella preparation (using the electron and ion beams of the dual beam system, respectively). The most remarkable Z-contrast is given by the insulating HfO$_2$ film and the surrounding layers. In this imaging method the brighter layers are responsible for incoherently scattering more electrons to the annular detector. The interface between the Co and the Pd appears drastically defined (darker than the surrounding) which might relate to the fact that Co was deposited by electron beam while Pd was sputtered \citep{zazpe_thesis}. Moreover, the Pd layer follows the topographic landscape of the outermost Co surface and no inter-diffusion can be clearly identified. On the contrary, the smooth transition from the HfO$_2$ to the Co layer seems to suggest that oxygen is present (or even HfO$_2$) within the Co layer, which could be explained in terms of its reactivity and scavenging ability \citep{khotseng_oxygen_2018}. 

As can be observed in Fig. \ref{fig:STEM}, specifically regarding the TiO$_{x}$ layer, a columnar arrangement of alternating brighter and darker areas is noticeable, which indicates inhomogeneous and localized oxidation. This new insight offered by the \textit{HAADF} image, in agreement with the previously mentioned oxygen presence registered along the whole thickness of the Ti layer \citep{quinteros_atomic_2018}, supports the hypothesis of a stack more complex than only a pure HfO$_2$ switching layer. Moreover, the apparent columnar structure determined by dark and bright zones within the TiO$_x$ layer may account for the quality of the former Ti layer, featuring a columnar growth with grain boundaries that could facilitate selective oxygen diffusion, that in turn could have lead to a laterally inhomogeneous oxidation all across the layer's thickness.   

It is also worth mentioning that even though the presence of oxygen along the whole Ti layer appears strongly supported in the experimental evidence (both by previously reported Secondary Ion Mass Spectroscopy \citep{quinteros_atomic_2018} as well as indirect quantification of the capacitive term \citep{quinteros_oxidos_2016} and the new insight offered here by the \textit{HAADF} measurement) neither of them enable an exact determination of the stoichiometry. A plethora of stable titanium oxides \citep{szot_tio2prototypical_2011} has been reported displaying a variety of dissimilar electrical behaviours. This, in addition to the mentioned inhomogeneity observed by the \textit{HAADF} image, impedes intentionally depositing a certain titanium oxide phase having to rely on following the same procedure flow such that the conditions are the appropriate ones to promote the oxidation of the Ti layer as it occurred in the original batch.     

\subsection*{Subsequent deposition processes}

Based on the aforementioned picture, the two subsequent deposition processes (batches B2 and B3) were focused on reproducing the TiO$_{x}$ and the HfO$_2$ layers. All the samples of the subsequent batches consist of a silicon highly-doped substrate, a silicon oxide layer, and a sputtered layer of Ti on top of which a layer of HfO$_2$ was deposited by \textit{ALD}. Top electrodes (\textit{TE}) of different sizes, shapes, and metals were defined. Except otherwise specified, all the deposited layers were continuous excluding the \textit{TE}, which were lithographically defined.   

B2 samples were aimed to investigate the effect of the Ti layer thickness\footnote{Within each batch, samples with different specifications were grown, resulting in labels such as B2 $\#$\textbf{i}.}, while keeping the thickness of the underlying SiO$_2$ layer within the same order of magnitude (further details in Table \ref{tab:FZJ-RUG-sample-details}). The used substrates were commercial ones and the doping level of the semiconductor assured both its metallic-like behaviour (to serve as a back contact) as well as the defective SiO$_{2-x}$ obtained as a consequence of thermal growth from a highly-doped buffer \citep{ho_si_1979}. The reactant used for the \textit{ALD} deposition differed from the original recipe in the use of water, instead of ozone, as oxygen precursor, in an attempt to decouple the two possible sources of oxygen in the Ti layer: the ozone during the \textit{ALD} reaction and the oxygen content in the HfO$_2$ and/or SiO$_{2-x}$ layers acting as reservoirs. 

\begin{figure}[ht!]
        \centering
        \begin{subfigure}[b]{0.47\textwidth}
        \centering
        \includegraphics[width=\textwidth]{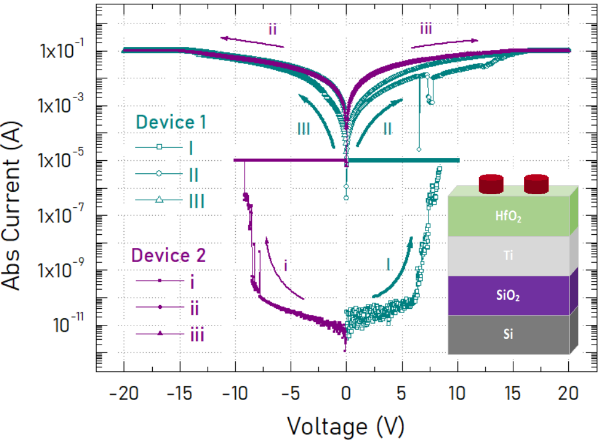}
		\subcaption{}        
		\label{fig:FZJ-IV}
        \end{subfigure}
        \hfill
        \begin{subfigure}[b]{0.47\textwidth}
        \centering
        \includegraphics[width=\textwidth]{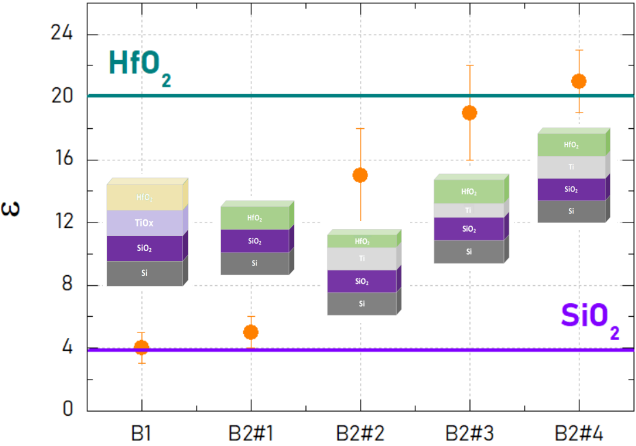}
		\subcaption{}  
		\label{fig:FZJ-dielectric}      
        \end{subfigure}
        \vspace{0.2cm}
        \caption{Graphical summary of B2 samples' electrical properties. \ref{fig:FZJ-IV} Current as a function of voltage (IV) in semilog scale. IV loops, starting towards opposite polarities for two equivalent pristine devices (1 and 2, respectively) of the same sample (B2$\#$4), are presented. A sketch of the stack is included (inset), demonstrating the continuity of the layers except for the \textit{TE}. The cartoon accounts for two different top electrodes defining two different devices within the same stack. \ref{fig:FZJ-dielectric} Dielectric constants determined from the normalized ratio $\frac{Capacitance}{Area}$ divided by the thickness of the dominant capacitor in each case. B2 samples, containing a Ti metallic-like layer, display a permittivity matching the HfO$_2$ value while the B1 sample (with a fully oxidized TiO$_x$ layer) matches the SiO$_2$ dielectric constant. The top electrodes are omitted to emphasize the differences among the stacks themselves.}
        \label{fig:FZJ-samples-summary}
\end{figure}

Fig. \ref{fig:FZJ-samples-summary} displays the electrical response of B2 samples, a sketch of which is included as an inset. Within this batch (B2) four different samples were grown: B2 $\#$1 (Si/SiO$_{2-x}$ (120 nm)/HfO$_2$ (20 nm)/Pt), B2 $\#$2 (Si/SiO$_{2-x}$ (120 nm)/Ti (20 nm)/HfO$_2$ (10 nm)/Pt), B2 $\#$3 (Si/SiO$_{2-x}$ (120 nm)/Ti (10 nm)/HfO$_2$ (20 nm)/Pt), and B2 $\#$4 (Si/SiO$_{2-x}$ (120 nm)/Ti (20 nm)/HfO$_2$ (20 nm)/Pt). Sample B2 $\#$4 is the most similar to B1 in terms of partial and total thicknesses of the laminated layers and is the one chosen for the devices in Fig. \ref{fig:FZJ-samples-summary}. On the one hand, three successive loops (labelled in Roman numbers) of two different pristine devices are shown. During a first forming-like curve (I), an extremely high-current regime is observed, even reaching the current compliance (10 $\mu$A) defined for preventing an irreversible breakdown\footnote{This is not strictly accurate since the compliance provided by the equipment relies on the switching of a relay which takes some instants to react during which the irreversible breakdown may occur anyway. Better alternatives consist of serially connecting physical devices that limit the current in a sustained way \citep{menghini_resistive_2015}.}. This regime is also observed after removing the power supply when stimulating towards the same polarity as in the previous run (II). Moreover, when reversing the polarity (III), the previously achieved low-resistance state is retained, opposite to the rectifying ability highlighted in the original case (B1). That occurs either if the first explored polarity is positive or negative, as demonstrated in Fig. \ref{fig:FZJ-IV} (successive I, II, and III and i, ii, and iii loops, for devices 1 and 2, respectively). Other samples of the same batch (B2) showed similar characteristics that, in turn, differ from B1 noticeably, mainly regarding the lack for self-limitation and the impossibility of recovering the pristine state.  

Furthermore, Fig. \ref{fig:FZJ-dielectric} compares the dielectric permittivity (measured at 10 kHz) obtained for different stacks, with their labels accounting for the various B2 samples (see Table \ref{tab:FZJ-RUG-sample-details}). An original B1 sample is included for the sake of comparison. The measured capacitance of B2 samples agrees with a dominant HfO$_2$-based capacitor for all the cases except for the sample without Ti layer. This can be rationalized as a persistently metallic Ti layer which determines two independent capacitors within the stack: a top one, composed by Pt/HfO$_2$/Ti whose area is determined by the top electrode's size, and a bottom one corresponding to Ti/SiO$_{2-x}$/Si whose area corresponds to the whole film area (due to devices' definition, see Fig. \ref{fig:FZJ-IV}'s inset).  

In summary, the electric behaviour of the B2 samples seems to indicate a simple bipolar switching scheme, instead of the striking properties observed in B1. Moreover, the capacitive measurements agree with retaining two serially-connected capacitive units decoupled from each other. Taken together, these two evidences are interpreted as an indication of the incomplete (if any) Ti oxidation. An aftermath analysis seemed to point out that the decision of mimicking the SiO$_{2-x}$ thickness 
could have been detrimental for reproducing the voltage drop distribution. An additional consideration corresponds to having used Pt in contact to the HfO$_2$ layer, which might have implied a 'less defective' HfO$_2$ layer (more stoichiometric and consequently less leaky) due to the lower reactivity of Pt compared to Co (as in B1). The conclusion that can be withdrawn is that the reactive environment within the \textit{ALD} chamber is key for the formation of the specific TiO$_x$ phase, as we have already suggested in a previous communication \citep{quinteros_atomic_2018}. This, in turn, is in agreement with the electrical behaviour found by Yoon and col. \citep{yoon_pt/ta2o5/hfo2x/ti_2015} who reported similar results on \textit{ALD}-grown HfO$_2$-based stacks using ozone as oxygen precursor\footnote{Their stack can be considered as half of the one introduced here as B1.}.  

A second attempt consisted of reproducing in more detail the HfO$_2$ deposition itself (batch B3). In this case, both the Hf and O precursors were the same as during the original process and the \textit{TE} material was also reproduced. The substrate was a highly-doped Si (same doping type) only covered by a native oxide layer as thin as $\sim$ 2 nm (no intentional growth of thermal SiO$_{2-x}$\footnote{This decision is partly supported by the assumption that the SiO$_2$ layer merely acts as a series resistor. Additionally, the structural properties of TiO$_x$ grown on chemically treated Si and thermally grown SiO$_2$ were demonstrated to be equivalent up to thicknesses of 20 nm \citep{puurunen_controlling_2011}.}). Apart from this last observation, the main difference between B1 and B3 is the temperature at which the \textit{ALD} was conducted. 

\begin{figure}[ht!]
        \centering
        \begin{subfigure}[b]{0.47\textwidth}
        \centering
        \includegraphics[width=\textwidth]{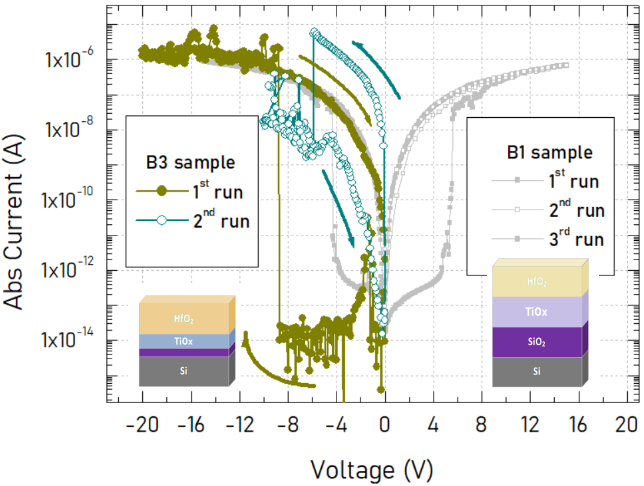}
		\subcaption{}        
		\label{fig:RUG-IV}
        \end{subfigure}
        \hfill
        \begin{subfigure}[b]{0.46\textwidth}
        \centering
        \includegraphics[width=\textwidth]{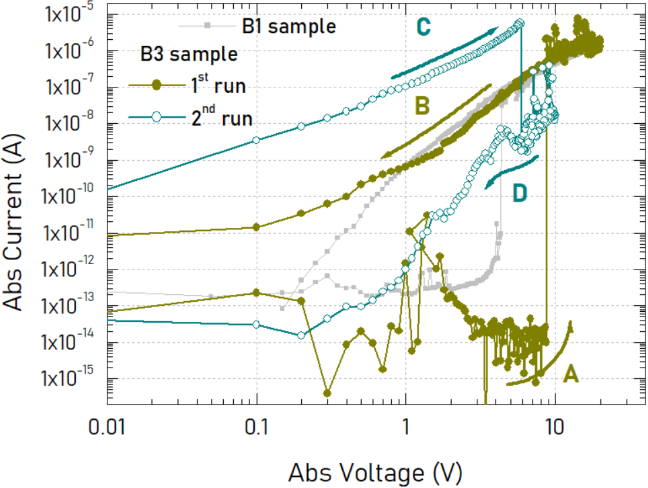}
		\subcaption{}  
		\label{fig:RUG-conduction}      
        \end{subfigure}
        \vspace{0.2cm}
        \caption{Electrical properties of a typical B3 sample. The comparison with a reference device (from a B1 sample) reflects a self-limiting mechanism at work in B3 samples, even when the SiO$_{2-x}$ thickness is reduced to its minimum. Furthermore, a metastable state (identified as branch B) partially matches the results obtained for the original batch. \ref{fig:RUG-IV} Current as a function of voltage (IV) in semilog scale of a B3 sample (coloured symbols) compared to the original device (grey scale). Successive IV loops are measured. \ref{fig:RUG-conduction} Double-logarithmic IV loop for the two successive excursions presented in Fig. \ref{fig:RUG-IV}. The dependence obtained for a sample of reference is included for comparison.}
        \label{fig:RUG-samples-summary}
\end{figure}

Fig. \ref{fig:RUG-samples-summary} includes IV loops recorded in a B3 sample consisting of the same nominal thickness of both Ti and HfO$_2$ with the \textit{ALD} process employing exactly the same precursors as in B1. Once again the electrical properties differ from those of B1 mainly in the absence of coupling between the two units, which gives rise to the rectifying ability, and in the lack of C2C and D2D stability (highlighted as a relevant and distinctive aspect of B1). Interestingly, like in the original case, there is a trace of a self-limiting switching that prevents the current from a runaway immediately after the SET operation is observed. However, this was not the case for all the samples of this batch. Fig. \ref{fig:RUG-conduction} shows that if there is any similarity between the two types of samples (see Fig. \ref{fig:RUG-conduction}) it is not stable enough to persist after cycling. Moreover, the voltage needed to SET the B3 devices is even higher than in B1. This value would even surpass the nominal breakdown field if the voltage would fully drop on the HfO$_2$ layer. Even though, this is not the case (no irreversible breakdown is observed), as demonstrated by the self-limited current, it does not appear as repeatable and sustainable as the process registered in B1. In addition, a second loop towards the same polarity starts at a resistance level lower than that at the end of the previous voltage run and evidences a RESET operation typical of an unipolar resistive switch (instead of the bipolar nature claimed in the rationalization of B1 electric behaviour).     

\begin{table}[h!]
    \centering
    \begin{tabular}{c|c|c|c}
     & B1 & B2 & B3 \\
    \hline
    Condition or parameter &  &  &  \\
    \hline
    Si doping ($cm^{-3}$) & $10^{18}/10^{19}$ (p-type) & $10^{18}$ \colorbox{red}{\color{white}{(n-type)}} &  \colorbox{green}{$10^{19}$ (p-type)}\\
    Oxidized SiO$_{2-x}$ thickness ($nm$) & 150 & 120
     & \colorbox{red}{\color{white}{0}} \\
    Ti thickness ($nm$) & 20 & 0, 10, \colorbox{green}{20} & 0, 5, 10, \colorbox{green}{17}, 35 \\
    Ti deposition method & sputtering & sputtering & sputtering \\
    ALD pre-process & - & - & \colorbox{red}{\color{white}{O$_3$ and N$_2$ pulses at 300C}} \\
    T during ALD ($C$) & 300 & \colorbox{red}{\color{white}{175}} & \colorbox{red}{\color{white}{100}} \\
    Oxygen precursor & ozone & \colorbox{red}{\color{white}{water}} & \colorbox{green}{ozone} \\
    Hf precursor & TDMAHf & \colorbox{red}{\color{white}{TEMAHf}} & \colorbox{green}{TDMAHf} \\
    HfO$_2$ thickness ($nm$) & 20 & 10, \colorbox{green}{20} & \colorbox{green}{20} \\
    TE material (thickness in $nm$) & Co(35)/Pd(40) &\colorbox{red}{\color{white}{Pt(70)}} & \colorbox{green}{Co(30)}/\colorbox{red}{\color{white}{Pt(40)}}, Pt(70), Ti(10)/Au(60) \\
    TE deposition method & e-beam/sputtering &  & \colorbox{red}{\color{white}{sputtering}} \\
    TE shape & square & circle & circle \\
    TE dimension ($\mu$m) & L = 200 & $\phi$ = 150, \colorbox{green}{250} & $\phi$ = 100, 125, 200, \colorbox{green}{250}, 350, 500, 1000 
    \end{tabular}
    \caption{Parameters' comparison between the original batch (B1) samples and the two subsequent deposition processes (B2 and B3) aiming to mimic the remarkable electrical properties of the former batch. TEMAHf stands for hafnium tetrakis (ethylmethylamide) = $Hf[N(CH_3)(C_2H_5)]_4$. TDMAHf stands for hafnium tetrakis (dimethylamide) = $Hf[N(CH_3)_2]_4$. Simiarities are highlighted in green while differences are highlighted in red.}
    \label{tab:FZJ-RUG-sample-details}
\end{table}

Table \ref{tab:FZJ-RUG-sample-details} summarizes \colorbox{green}{similarities} (highlighted in green) and \colorbox{red}{\color{white}{differences}} (highlighted in red) compared to the original process. Unfortunately, neither the batch B2 nor the B3 one gave the expected results. Even though it remains unclear whether the accidentally formed TiO$_x$ phase from the original samples was matched in the later ones, the fact that the most remarkable properties were not reproduced make us consider other possibilities.

\section*{Discussion}

Even though the space of parameters of such a complex stack is far from fully explored, the evident contrast between the remarkable features of the original samples and the electrical response of the subsequently-designed ones led us to wonder if there is not any other suitable description for the observed behaviour. Perhaps the key for unlocking the desired properties lies on an alternative physical mechanism at work, while the previous picture, based on the role of the TiO$_x$ interface, might have been misleading the efforts in that regard. In the search for other explanations, in the following, we discuss two alternatives that seem feasible and deserve further investigation: threshold switching and (anti)ferroelectricity, including a mention to ferroelectricity for the sake of completeness.  

\subsection*{Threshold switching}  

\begin{figure}[ht!]
       \centering
       \begin{subfigure}[b]{0.45\textwidth}
       \centering
       \includegraphics[width=\textwidth]{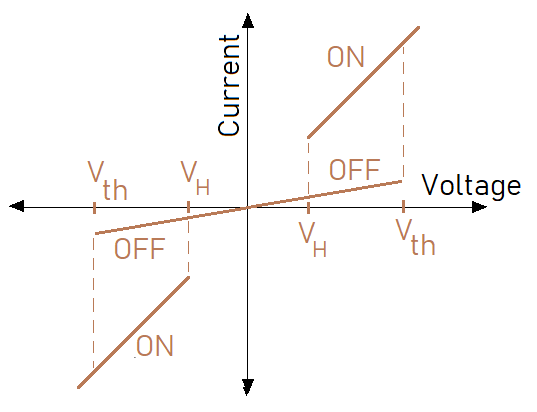}
		\subcaption{}        
		\label{fig:Typical-threshold}
        \end{subfigure}
        \hfill
        \begin{subfigure}[b]{0.45\textwidth}
        \centering
        \includegraphics[width=\textwidth]{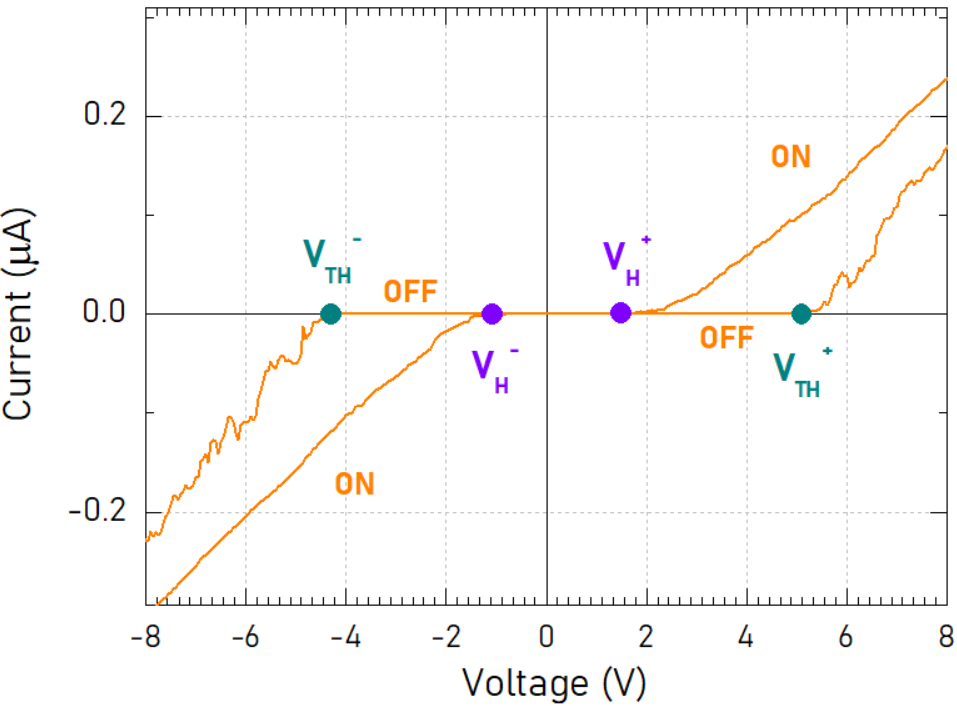}
		\subcaption{}  
		\label{fig:Cont-LinIV}      
        \end{subfigure}
        \vspace{0.2cm}
        \caption{\ref{fig:Typical-threshold} Typical threshold switching IV dependence. Adapted from Adler \textit{et al.} \citep{adler_mechanism_1978}. The device naturally starts at the OFF state that upon V$_{th}$ undergoes a transition to an ON state retained only until V$_h$ is surpassed. \ref{fig:Cont-LinIV}
Current measured on a B1 sample upon a continuous voltage sweep. The \textit{LRS} vanishes before removing the power supply.}
        \label{fig:threshold-switching}
\end{figure}

In the early '60s, the observation of a sudden change in the resistance state in bulky chalcogenides was referred to as threshold switching \citep{ovshinsky_reversible_1968}. In those systems, when a certain threshold voltage value (V$_{th}$) is reached\footnote{That value depends on the thickness of the material.}, a dramatic reduction of the resistance defines a new state that will be held until the so-called holding voltage is reached back (see Fig. \ref{fig:Typical-threshold}). This second voltage condition (V$_h$) is obtained while reducing the applied stimulus since it is lower than the threshold one (V$_{th}$), of the same polarity. In such a case, when the device is not being polarized, the highly resistive state is recovered. The same qualitative behaviour is observed regardless the polarity. An archetypical electrical response is presented in Fig. \ref{fig:Typical-threshold} (adapted from ref. \citep{adler_mechanism_1978}). 

Fig. \ref{fig:Cont-LinIV} displays an IV loop conveniently presented to analyse differences and similarities with the threshold switching. The loop was recorded in a B1 sample and acquired under a continuous voltage ramp\footnote{A device measured on the same B1 sample was shown for the sake of comparison in Fig. \ref{fig:RUG-IV}. The differential electrical response (non-volatile in that case) relies on the difference in the applied protocol as highlighted in the subsection \textit{Original batch}.}. Under such conditions, the volatile switching nature is evident since the ON state vanishes before returning to the 0 V condition. Presented in this way, a threshold could be spotted even though the switch is not as abrupt as in the archetypical case (Fig. \ref{fig:Typical-threshold}). In addition, an asymmetry between the positive and the negative polarities is observed which could screen some other relevant aspect not taken into account. The fact that the original samples (B1) appear so dependent on the specific pulsing scheme seems to suggest a strong dependence either on thermal activity or transient responses which would enable different responses depending on the waiting time in between pulsing. 
 
Even when there might be a reminiscence of the electrical signature of the original threshold switches, the nanometric stacks are radically different from the glassy semiconductors in which the effect was firstly identified. Nowadays, they are composed by nanolaminated layers of binary or ternary oxides. For contemporary systems like these, for instance a Poole-Frenkel barrier lowering (field driven effect) and an associated thermal runaway have been proved to satisfactory reproduce experimental results in niobium oxides \citep{funck_multidimensional_2016}. In fact, quite recently, the debate between thermal and non-thermal mechanisms has been stirred up due to an experimental demonstration of the heat dissipated during a CC-NDR experiment \citep{goodwill_spontaneous_2019}. 

\vspace{0.2 cm}

Therefore, on the one hand, the threshold nature of the electrical response offers a framework to rationalize the observed changes in thick layers without any requirement for specific crystal structures, as is the case for the samples presented in this study. It would also allow to rationalize the occurrence of volatile switching. Eventually, the coexistence of volatile and non-volatile states might also be explained in terms of the dissipated heat due to the specific applied protocol. On the other hand, this picture considers only two relevant interfaces while our stacks are nanolaminated or multi-layered by construction. In order to further test if this mechanism is at play it would be necessary to identify which are those and what role the others have. In addition, the similarity in the SET voltages for the two polarities should be explained in terms of a stack that does not present any symmetry.  

\subsection*{Ferroelectric HfO$_2$-based systems}

Ferroic materials are characterized by a spontaneous order parameter that can be reversibly switched between at least two energetically-equivalent ground states by an applied conjugate field. \textit{Ferroelectrics} (\textit{FE}) are insulators possessing a spontaneous electric polarization switchable by an electric field \citep{rabe_physics_2007}. 

Recently, ferroelectricity has been observed in crystalline doped HfO$_2$. The most interesting aspect of this ferroelectricity\footnote{Understood as the presence of a remnant polarization able to be switched upon external stimuli.} is that, contrary to the expectations of losing the polarization while the thickness is reduced, hafnia seems to display stronger polarization the thinner the layer becomes. And even though there are many nuances still under investigation, the surface energy contribution and associated internal pressure inside the nano-size grains seems to be the reason for this non-intuitive behaviour \citep{martin_ferroelectricity_2014,sang_structural_2015}.  

Within this context, one could wonder whether there is any chance that the HfO$_2$ layer formed in our stack also presents these characteristics. Even though the process temperature should not be high enough to crystallise hafnia, it shows a polycrystalline nature \citep{zazpe_resistive_2013}, while usual \textit{ALD} processes give rise to amorphous layers, in need for a \textit{rapid thermal anneal} (\textit{RTA}) to crystallize. In this regard, it is worth bearing in mind that doping, in the form of migrating ions from layer to layer, could also affect this aspect by introducing distortion in the lattice. Even though one might guess whether there is any possibility for the Pd to have diffused into the HfO$_2$ layer \citep{chakraborty_mechanisms_2010} to introduce that kind of crystalline defect, the STEM measurement rules out this possibility, given the sharply defined Co/Pd imaged interface. In addition, the thickness seems to also play a role in triggering the crystallization \citep{polakowski_ferroelectricity_2015}. In any case, even if we accept the evidence of the polycrystallinity of the layer, grazing-incidence X-ray measurements suggested a dominant monoclinic phase \citep{zazpe_resistive_2013} which being centro-symmetric, would be incompatible with the presence of ferroelectricity. Even though a refinement of the identified peaks might be necessary, there is no evidence for a ferroelectric behaviour either from the structural point of view or from electrical switching measurements (not shown) using the so-called PUND protocol \citep{schenk_about_2014}.

\vspace{0.2 cm}
Having concluded that none of the signatures seem to indicate a ferroelectric nature of our stack, a closer look at the I-V loop allows to recognize a striking similarity with the polarization vs field (P-E) dependence of an antiferroelectric (\textit{AFE}). One may wonder whether there is any mechanism to relate the current to the polarization such that the IV curve resembles so closely the dependence of the strain upon the same driving force.

\subsection*{Antiferroelectric HfO$_2$} 

\begin{figure}[ht!]
        \centering
        \begin{subfigure}[b]{0.45\textwidth}
        \centering
        \includegraphics[width=0.9\textwidth]{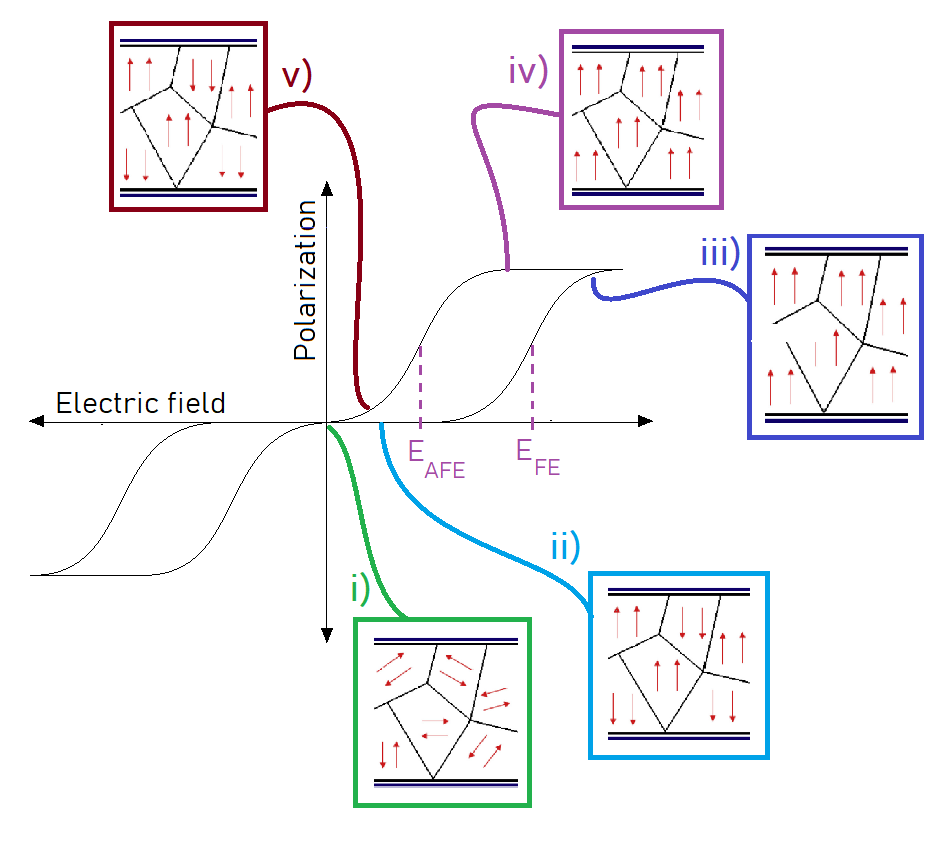}
		\subcaption{}        
		\label{fig:typicalAFEloop}
        \end{subfigure}
        \hfill
        \begin{subfigure}[b]{0.45\textwidth}
        \centering
        \includegraphics[width=\textwidth]{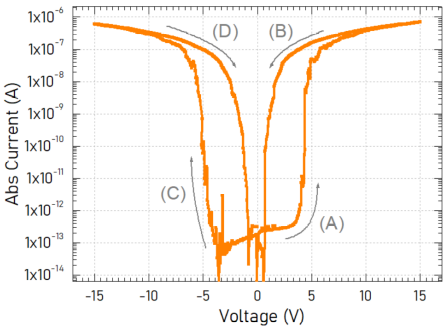}
		\subcaption{}  
		\label{fig:nGlinIV-similAFE}      
        \end{subfigure}
        \vspace{0.2cm}
        \caption{Antiferroelectric behaviour compared to experimental data obtained in a B1 device. \ref{fig:typicalAFEloop} Archetypical polarization vs electric field (P-E) loop for an \textit{AFE} with cartoons of ferroelectric domains' orientation upon external electric field representing the \textit{AFE/FE} transition (adapted from Hao \textit{et al.} \citep{hao_comprehensive_2014}). \ref{fig:nGlinIV-similAFE} Semi-logarithmic plot of current (absolute value) vs voltage measured upon a continuous voltage ramp (same curve as displayed in Fig. \ref{fig:Cont-LinIV}).}
        \label{fig:AFE_P-V_loop}
\end{figure}

The P-E loop for an \textit{AFE} material can be described as a two threshold excursion \citep{kittel_theory_1951}. Firstly, upon increasing the applied electric field a pronounced increase of the polarization accompanied by a transient current is observed, very much like a \textit{FE} itself \citep{schenk_about_2014}. That condition corresponds to the re-orientation of the domains that were antiparallel to the external field direction. The \textit{FE} state of \textit{AFE} materials is however, unstable and depending on their ability to hold this state they are classified as soft or hard \textit{AFE} \citep{hao_comprehensive_2014}. In turn, this characteristic would depend on thermal dissipation, strain and other factors (such as charge rearrangement, as shown in \textit{AFE}-like loops that upon stimulating undergo a transition to pure \textit{FE} nature, also known as wake-up effect \citep{zhou_wake-up_2013}). Most of the \textit{AFE} show a distinctive feature while the applied electric field is reduced, at a certain value (E$_{AFE}$ in Fig. \ref{fig:typicalAFEloop}) the external stimulus will not be enough to hold the parallel polarization and the material will relax to its more energetically stable \textit{AFE} state. An abrupt reduction of the polarization (without considering the paraelectric component, it would ideally reach zero) accompanied by a current flow in the direction of the field gradient \citep{schenk_about_2014} would mark the \textit{FE}/\textit{AFE} transition. Ideally, at zero bias, there is no trace of the phase transition.

Even when \textit{FE} and \textit{AFE} definitions are different and so are the requirements for the materials able to display those behaviours \citep{rabe_antiferroelectricity_2013}, there is a close relationship between the two of them. The energies of the \textit{FE} and \textit{AFE} phases are supposedly very close together being able to transition from one another upon varying different parameters. In fact, \textit{FE} behaviour of thin films of otherwise \textit{AFE} materials was demonstrated \citep{ayyub_ferroelectric_1998} while tuning the dopant content allows to cover many possibilities in a complex phase diagram of commensurate and incommensurate \textit{FE} \citep{asada_coexistence_2004} as well as ferrielectricity and \textit{AFE} \citep{asada_coexistence_2004}. 

Therefore, given that ferroelectricity was identified in HfO$_2$-based systems, one may wonder if \textit{AFE} could take place upon a proper orientation. Hafnia based systems have broadly demonstrated \textit{FE} properties attributed to its intrinsic nature but achieved by stabilizing otherwise unstable phases \citep{polakowski_ferroelectricity_2015}. Even though a polar distortion of a high symmetry phase is needed for a \textit{FE} to occur, pinched hysteresis loops as well as the wake-up and fatigue are still being closely studied \citep{pesic_physical_2016}, and hysteresis deformation in general \citep{schenk_about_2014} suggests subtle acting mechanisms that could lead to extremely interesting behaviours still to be detected. In particular, quite recently the pinched loop during the so-called wake up of \textit{FE} hafnia was compared to the double loop typical of the \textit{AFE} response \citep{lomenzo_depolarization_2020}.  

\vspace{0.2 cm}
As a summary, it is now well established that \textit{FE} in HfO$_2$-based systems can be achieved by different methods, either by doping, encapsulating or both, such that metastable phases of lower symmetry can be stabilized. 
Epitaxial strain can also stabilize the \textit{FE} phase but it is incompatible with our system. In addition, \textit{AFE} in hafnia-based materials have also been reported. Our system of interest, while possibly displaying some fraction of \textit{FE} phase, resembles to some extent the behaviour of an \textit{AFE} (Fig. \ref{fig:nGlinIV-similAFE} coupled to an additional mechanism that is able to map polarization into current). Therefore, the next step will be to explore the possibility of having a ferroic hafnia which adequately coupled to other layers results in the occurrence of alternative volatile and non-volatile states of high interest for improved functionality devices with neuromorphic scopes.   

\section*{Final remarks and perspectives}

There exist several differences among samples grown in different set-ups. Assuming that the SiO$_{2-x}$ layer does not play a key role in the switching mechanism, the effort was focused on mimicking the TiO$_x$ and HfO$_2$ quality. Also, the role of the top electrode was somewhat disregarded. Nevertheless, the impossibility of reproducing the combination of properties featured by B1 indicated a key aspect remained undetected.  

Understanding the ubiquity of threshold switching and the variety of underlying processes leading to similar macroscopic response, it seems clear that charge injection could be at play in our stack. This, in particular, could help identifying the mechanism that enables the observed volatile switching. 

In addition, having determined the \textit{FE} nature of many hafnia-based systems leads us to consider the plausibility of an unintentional ion migration that would populate the layer, introducing the necessary distortion to form lower symmetry phases. Nonetheless, the impossibility of recording a \textit{FE} loop and the lack of evidence for a predominant structural phase contradicts this scenario. However, given the striking similarities between the IV loop and the displacement in an \textit{AFE}, the hypothesis of this connection appears unavoidable. 

If anything like a ferroic order of the HfO$_2$ layer is involved (\textit{FE} or \textit{AFE}), then the temperature inside the \textit{ALD} chamber during deposition would be of paramount importance. So far, for the subsequent batches (B2 and B3), the temperature was held high (at 300$^{\circ}$C) only for a couple of minutes before the introduction of the HfO$_2$ precursors, to promote the oxidation of the underlying layer. Depositing HfO$_2$ at such temperatures had been, until now, intentionally avoided to preserve the quality of the precursors. Performing an ex-situ \textit{RTA} would enable to crystallize the HfO$_2$ (which in batches B2 and B3 seem to be amorphous) but would not produce the same impact on the Ti layer underneath.  

An unexplored aspect of the stack is the role of the Co layer used as top electrode in the original batch. Cobalt is ferromagnetic (\textit{FM}) with the magnetization out-of-plane when the layer is $\leq$ 1.2 nm and in-plane otherwise \citep{metaxas_creep_2007}. In addition, CoO$_x$, which could have been formed (as suggested from the \textit{HAADF} image) due to the oxygen scavenging from the oxide matrix, is antiferromagnetic. Any of these scenarios could impact the overall properties of the stack and have not been extensively studied. In fact, deposition of \textit{FM} layers on top of \textit{FE} dielectrics has been a strategy exploited to achieve electrical control of \textit{FM} loops \citep{vermeulen_ferroelectric_2019}.

Another unconsidered ingredient is the role of the oxygen vacancies. Even though, we cannot rule out the presence of oxygen vacancies within any of our stacks, experimental evidence seems to indicate other mechanism/s controlling the overall response. For that reason, and given the fact that involving oxygen vacancies would have required a detailed analysis on its own, we have avoided involving them in our proposed alternative explanations.  

\vspace{0.2cm}
 
In summary, a list of changes to be applied to the fabrication flow as well as a complementary characterization have been identified as necessary in order to spot traces of the mentioned mechanisms. We hope that this discussion will encourage others to join efforts towards a tight control of the stack in the direction of obtaining the desired properties. 

\vspace{6pt} 



\section*{Acknowledgements}


\noindent The authors would like to acknowledge R. Zazpe and L. Hueso for providing the samples that triggered this project. Special thanks has to be given to A. Hardtdegen and S. Hoffmann-Eifert who were involved in the deposition of the \textit{FZJ} complementary batch. 

\vspace{0.2cm}

\noindent The authors gratefully acknowledge financial support from NWO’s TOP-PUNT grant 718.016002. C.P. Quinteros also wants to acknowledge \textit{Deutscher Akademischer Austauschdienst} (\textit{DAAD}) and the Argentinian Education Ministry for funding her stay at the \textit{Forschungszentrum Jülich}. This project has also received funding from EU-H2020-RISE project \textit{Memristive and multiferroic materials for logic units in nanoelectronics} 'MELON' (SEP-2106565560).  

%



\section*{Conflicts of interest} 

The authors declare no conflict of interest. 

\bibliographystyle{ieeetr}
\bibliography{LibNov27}





\end{document}